\documentclass[runningheads]{llncs}

\usepackage{cite}
\usepackage{hyperref}
\usepackage{bm}
\usepackage{amsmath}
\usepackage{tabularx}
\usepackage[percent]{overpic}
\usepackage{fontawesome}
\usepackage{color,xcolor}
\usepackage[capitalize, nameinlink]{cleveref}
\crefname{section}{Section}{Sections}
\Crefname{section}{Section}{Sections}
\crefname{figure}{Fig.}{Figs.}
\Crefname{figure}{Fig.}{Figs.}
\crefname{table}{Table}{Tables}
\Crefname{table}{Table}{Tables}

\newcommand{\ignore}[1]{}

\begin{document}
\title{Predicting Clinical Outcome of Stroke Patients with Tractographic Feature}

\author{Po-Yu Kao\inst{1}\orcidID{0000-0002-9439-8819} \and
Jeffereson W. Chen\inst{2} \and
B.S. Manjunath\inst{1}}
\authorrunning{P.-Y. Kao et al.}

\institute{University of California, Santa Barbara, California, United States \\ \email{\{poyu\_kao,manj\}@ucsb.edu} \and
University of California, Irvine, California, United States
}

\maketitle
\setcounter{footnote}{0}
\begin{abstract}
The volume of stroke lesion is the gold standard for predicting the clinical outcome of stroke patients. However, the presence of stroke lesion may cause neural disruptions to other brain regions, and these potentially damaged regions may affect the clinical outcome of stroke patients. In this paper, we introduce the tractographic feature to capture these potentially damaged regions and predict the modified Rankin Scale (mRS), which is a widely used outcome measure in stroke clinical trials. The tractographic feature is built from the stroke lesion and average connectome information from a group of normal subjects. The tractographic feature takes into account different functional regions that may be affected by the stroke, thus complementing the commonly used stroke volume features. The proposed tractographic feature is tested on a public stroke benchmark Ischemic Stroke Lesion Segmentation 2017 and achieves higher accuracy than the stroke volume and the state-of-the-art feature on predicting the mRS grades of stroke patients. In addition, the tractographic feature also yields a lower average absolute error than the commonly used stroke volume feature. 

\end{abstract}
\section{Introduction}
According to the World Health Organization, 15 million people suffer strokes each year, the second leading cause of death (5.8 million) and the third leading cause of disability worldwide \cite{johnson2016stroke,lindley2018stroke}. Around 87\% of strokes are ischemic strokes, which result from an obstruction within a blood vessel in the brain \cite{summers2009comprehensive}.
The corresponding lack of oxygen results in different degrees of disability of people, and the modified Rankin Scale (mRS) is commonly used to measure the degree of disability or dependence in the daily activities of stroke patients \cite{banks2007outcomes,goyal2016endovascular,van1988interobserver}. 

Several studies \cite{banks2007outcomes,lev2001utility,louvbld1997ischemic,parsons2000combined,van1998diffusion,vogt2012initial} demonstrate significant correlations between stroke volume and mRS grades, with larger lesions predicting more severe disability. 
However, only a few studies \cite{choi2016ensemble,forkert2015multiclass,maier2016predicting} extracted different features, including first-order features and deep features, other than volume from stroke lesion to predict the mRS grades of stroke patients.
The study of Maier and Handels \cite{maier2016predicting} is most relevant to our work. 
They extracted 1650 image features and 12 shape characteristics from the stroke volume, the volume surrounding the stroke and the remaining brain volume, and they applied a random forest regressor with 200 trees on these 1662 features to predict the mRS grades of stroke patients. 
However, the presence of stroke lesion may disrupt other brain regions that may affect the clinical outcome of stroke patients.

The main contribution of this paper is the introduction of a new second-order feature, the tractographic feature, that couples the stroke lesion of a patient with the average connectome information from a group of normal subjects.
The tractographic feature describes the potentially damaged brain regions due to the neural disruptions of the stroke lesion. 
Ideally one would like to use the diffusion images from the stroke patient, but this is not a realistic scenario. 
For instance, the patient with mental in their body is unsafe for getting an MRI scan. 
Instead, we use the ``normal'' subject data from the HCP project with the assumption that the parcellations and the associated tracts computed from that data are a reasonable approximation to extract the connectivity features. 
These tractographic features coupled with the stroke lesion information are used to predict the mRS grades of stroke patients.  
The concept of the tractographic feature was first proposed by Kao et al. \cite{kao2018brain} who used these to predict the overall survival of brain tumor patients. 
We modify their method to adapt to the size of the lesions and propose a new weighted vector of the tractographic feature. 
Our experimental results demonstrate that the proposed approach improves upon the state-of-the-art method and the gold standard in predicting the clinical outcome of stroke patients. 

\section{Materials and Methods}
\subsection{Dataset}
%
Ischemic Stroke Lesion Segmentation (ISLES) 2017 \cite{kistler2013virtual,maier2017isles} provides 43 subjects in the training dataset. Each subject has two diffusion maps (DWI, ADC), five perfusion maps (CBV, CBF, MTT, TTP, Tmax), one ground-truth lesion mask and clinical parameters. The ground-truth lesion mask is built in the follow-up anatomical sequence (T2w or FLAIR) and the corresponding mRS grade was given on the same day. The clinical parameters include mRS grade ranging from 0 to 4, time-to-mRS (88 to 389 days), TICI scale grade from 0 to 3, time-since-stroke (in minutes), and time-to-treatment (in minutes). Since TICI scale grade, time-since-stroke, and time-to-treatment were missing for some subjects, these three clinical parameters are not used in this work. The dimension and voxel spacing of MR images are different between each subject, but they are the same within each subject. We only focus on the subjects who obtain an mRS grade at 3 months (90 days) following hospital discharge since ascertainment of disability at 3-month post-stroke is an essential component of outcome assessment in stroke patients \cite{goyal2016endovascular}, and the tractographic data may change at a different time. Therefore, only 37 subjects are considered in this paper. 
\subsection{Tractographic Feature}
\begin{figure*}[htbp]
    \centering
    \includegraphics[width=\textwidth]{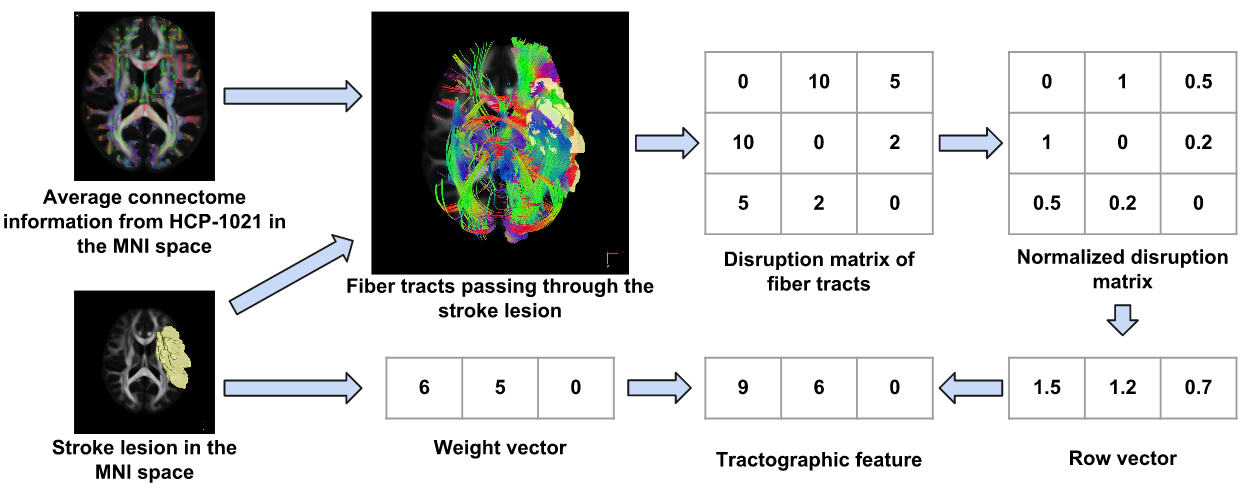}
    \caption{The workflow for constructing a tractographic feature from a stroke region.}
    \label{fig:tractographic_features_workflow}
\end{figure*}

The tractographic feature describes the potentially damaged region impacted by the presence of the stroke lesion through the average connectome information from 1021 Human Connectome Project (HCP) subjects \cite{van2013wu}. For each HCP subject, q-space diffeomorphic reconstruction \cite{yeh2011ntu} is used to compute the diffusion orientation distribution function. \cref{fig:tractographic_features_workflow} shows the workflow of building a tractographic feature for a stroke patient. 

Given the stroke lesion in the subject space, we first map the stroke lesion to the Montreal Neurological Institute (MNI) space \cite{grabner2006symmetric}. Second, we place one seed within each voxel of the brain region, and a deterministic diffusion fiber tracking method \cite{yeh2013deterministic} is used to find all possible tracts passing through the stroke volume inside the brain from the average diffusion orientation distribution function of 1021 HCP subjects. Topology-informed pruning \cite{yeh2019automatic} is used to remove false-positive tracts. 
Third, an existing brain parcellation atlas is used to create a disruption matrix $D$, which describes the degree of disruption between different brain parcellation regions due to the presence of the stroke lesion.

\begin{equation} \label{eq:disruption}
D = 
\begin{bmatrix}
d_{11}       & d_{12} & \dots & d_{1N} \\
d_{21}       & d_{22} & \dots & d_{2N} \\
\vdots       & \vdots & \ddots& \vdots \\
d_{N1}       & d_{N2} & \dots & d_{NN}
\end{bmatrix}
\end{equation}
$d_{ij}$ notes the number of tracts starting from a region $i$ and ending in a region $j$, and $N$ is the total number of brain parcellation regions in the existing atlas. 
Then, this disruption matrix is normalized by its maximum value, i.e., $\hat{D} = D/d_{m}$ where $\hat{D}$ is the normalized disruption matrix, and $d_{m}$ is the maximum element of the disruption $D$. 
Afterward, we sum up each column in this normalized disruption matrix $\hat{D}$ to form a row vector $\vec{L} = \sum_{i=1}^{N} \hat{d}_{ij} = [l_1, l_2, \dots, l_N] $. From the stroke lesion, we build a weight vector $\vec{\gamma}=[s_1, s_2, \dots, s_N]$, which is the distribution of the stroke volume in the different brain parcellation regions. $s_i$ is the volume of the stroke lesion in the $i$-th brain parcellation region. In the end, the row vector $\vec{L}$ is multiplied by this weight vector $\vec{\gamma}$ element-wisely to form the tractographic feature $\vec{T}$. 
\begin{equation} \label{eq:weighted_tf}
\vec{T} 
= \vec{\gamma} \circ \vec{L} 
\end{equation}
$\circ$ is the Hadamard-product. This vector $\vec{T}$ is the proposed tractographic feature extracted from stroke lesion without any diffusion information of a patient. In this paper, the Automated Anatomical Labeling (AAL) \cite{tzourio2002automated} template is used to define 116 brain regions so the dimension of the tractographic feature is 116. The reasons for choosing AAL rather than other existing atlases are (i) this atlas contains an optimal number of brain regions that could make each region large enough to compensate possible stroke-induced lesion effect or distortion, and (ii) this atlas contains cortical, subcortical and cerebellar regions, which could be equally important for mRS prediction. 
The source code is available on GitHub\footnote{\url{https://github.com/pykao/ISLES2017-mRS-prediction}}.

\subsection*{Parameters of Fiber Tracking} \label{section:fiber_parameters}

DSI Studio\footnote{\url{https://github.com/frankyeh/DSI-Studio}} is used to build the fiber tracts for each subject. 
\cref{tab:tracking} shows the tracking parameters\footnote{parameter\_id=F168233E9A99193F32318D24ba3Fba3Fb404b0FA43D21D22cb01ba02a01d} we used in this paper. 
The type of the stroke lesion is set to ROI (--roi=stroke\_lesion) that found all possible tracts passing through the stroke lesion.

\begin{table*}[htbp]
    \centering
    \caption{Tracking parameters of building the fiber tracts for stroke patients in this paper. More details of parameters can be found at \url{http://dsi-studio.labsolver.org/Manual/Fiber-Tracking}.}
    \begin{tabular}{c c}
        \hline
        \textbf{Parameter} & \textbf{Value} \\\hline
        Termination Index & qa \\ 
        Threshold & 0.15958 \\ 
        Angular Threshold & 90 \\ 
        Step Size (mm) & 0.50 \\ 
        Smoothing & 0.50 \\ 
        Min Length (mm) & 3.0 \\ 
        Max Length (mm) & 500.0 \\ 
        Topology-Informed Pruning (iteration) &  1 \\ 
        Seed Orientation & All orientations \\ 
        Seed Position & Voxel \\ 
        Randomize Seeding & Off \\ 
        Check Ending & Off \\ 
        Direction Interpolation & Tri-linear \\ 
        Tracking Algorithm & Streamline(Euler) \\
        Terminate if & 2,235,858 Tracts \\ 
        Default Otus & 0.60 \\ \hline
    \end{tabular}
    \label{tab:tracking}
\end{table*}

\subsection*{Parameters of Connectivity Matrix} \label{section:con_matrix_parameters}
DSI studio is used to create the connectivity matrix \footnote{\url{http://dsi-studio.labsolver.org/Manual/command-line-for-dsi-studio}} followed by fiber tracking. 
Automated Anatomical Labeling is chosen to form a $116\times116$ connectivity matrix.
The type of the connectivity matrix is set to end , the value of each element in the connectivity matrix is the count of fiber tracts , and the threshold to remove the noise in the connectivity matrix is set to 0 .

\subsection{Evaluation Metrics}
The employed evaluation metrics are (i) the accuracy, which is the percentage of the predicted mRS scores matching the corresponding ground-truth mRS scores, and (ii) the average absolute error between the predicted mRS scores and the corresponding ground-truth mRS scores. 
\section{Experimental Results}

\subsubsection{First experiment:}
\begin{figure*}[htbp]
    \centering
    \includegraphics[width=\textwidth]{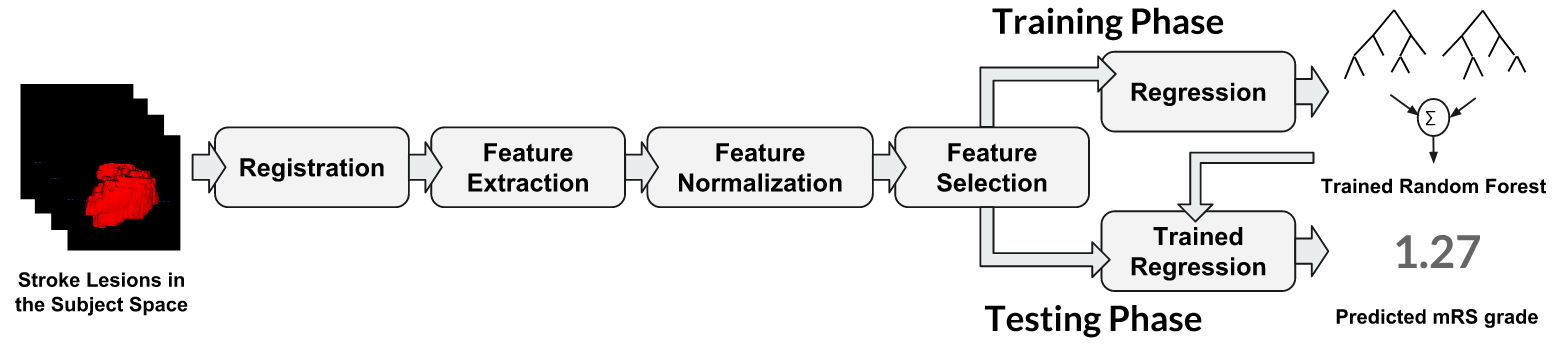}
    \caption{The framework of the first experiment. In the end, the predicted mRS grade is rounded to an integer.}
    \label{fig:pipe_exp_1}
\end{figure*}

In this experiment, we compare the mRS prediction performance of the tractographic feature with other first-order features extracted from the lesion mask. These first-order features include the volumetric feature, spatial feature, morphological feature and volumetric-spatial feature depicted in \cref{tab:first_order_features}. The framework of the first experiment is shown in \cref{fig:pipe_exp_1}. 

We first register the stroke lesions from subject space to the MNI space in order to overcome the differences of the voxel spacing and image dimension between different subjects. The tractographic feature and other first-order features are extracted from these normalized stroke lesions.  After feature extraction, we apply a standard feature normalization on the extracted features to ensure that each dimension of the features has the same scale. Then, we remove the dimensions of the features with zero variance between subjects and apply a recursive feature elimination with leave-one-out cross-validation to find the best subset of feature that yields the lowest average mean absolute error. In the training phase, we train one random forest regressor for each type of feature, i.e., five random forest regressors are trained. Each random forest regressor has 300 trees of which maximum depth is 3. In the testing phase, we use different types of features with the corresponding trained random forest regressors to predict the mRS grades of stroke patients, and the predicted mRS grade is rounded to an integer. We evaluate the mRS prediction performance of different types of features with leave-one-out cross-validation on ISLES 2017 training dataset. The quantitative results are reported in \cref{tab:features}. From \cref{tab:features}, the tractographic feature has the highest accuracy and lowest average absolute error on predicting the mRS grades of stroke patients compared to other first-order features. 

\begin{table*}[htbp]
    \centering
    \caption{First-order features extracted from the stroke lesion.}
    \begin{tabularx}{\textwidth}{c X}\hline
        \textbf{Type of feature} & \textbf{Descriptions} \\\hline
        Volumetric feature & Volumetric feature is the volume of the lesion in the MNI space, and it only has one dimension. \\
        Spatial feature & Spatial feature describes the location of the lesion in the brain. The centroid of the lesion is extracted as the spatial feature for each subject, and the spatial feature has three dimensions. \\
        Morphological feature & Morphological feature describes shape information of the lesion. The length of the major axis and minor axis of the lesion, the ratio of the length of major axis and minor axis of the lesion, the solidity and roundness of the lesion, and the surface of the lesion are extracted as the morphological feature. The morphological feature has six dimensions for each subject. \\
        Volumetric-spatial feature & Volumetric-spatial feature describes the distribution of the stroke lesion in different brain parcellation regions from an existing structural atlas. Automated Anatomical Labeling (AAL) \cite{tzourio2002automated} is used to build the volumetric-spatial feature so the dimension of the volumetric-spatial feature is 116. \\\hline
    \end{tabularx}
    \label{tab:first_order_features}
\end{table*}

\subsubsection{Second experiment:} 
We compare the mRS prediction performance of the tractographic feature with the state-of-the-art feature proposed by Maier and Handels \cite{maier2016predicting}.  We implement their feature extraction method on ISLES 2017 dataset.
First, 1650 image features and 12 shape features are extracted from the lesion volume and the apparent diffusion coefficients (ADC) maps in the subject space. Thereafter, these two types of features are concatenated to build a 1662-dimension feature. Then, we apply the same feature normalization, feature selection, cross-validation, and random forest regressor as the first experiment to predict the mRS of stroke patients. 
The quantitative results of the state-of-the-art feature are also shown in \cref{tab:features}. From \cref{tab:features}, the tractographic feature also achieves higher accuracy and similar average absolute error ($p=0.81$) compared to the state-of-the-art feature.

\begin{table*}[htbp]
    \centering
    \caption{The mRS prediction performance of different types of features on ISLES 2017 training dataset with leave-one-out cross-validation. The bold numbers show the best performance.  (The average absolute error is reported as mean $\pm$ std.)}
    \begin{tabular}{c  c  c } 
    \hline
    \textbf{Type of feature} & \textbf{Accuracy} & \textbf{Average absolute error} \\ 
    \hline
    Tractographic feature & \textbf{0.622} & $0.487 \pm 0.683$  \\ 
    Volumetric feature & 0.514 & 0.595 $\pm$ 0.715  \\ 
    Volumetric-spatial feature & 0.568 & 0.621 $\pm$ 0.817 \\ 
    Morphological feature & 0.378 & 0.703 $\pm$ 0.609 \\ 
    Spatial feature & 0.351 & 0.919 $\pm$ 0.882 \\ 
    Maier and Handels \cite{maier2016predicting} & 0.595 & \bm{$0.460 \pm 0.597$} \\ \hline
    \end{tabular}
    \label{tab:features}
\end{table*}

\section{Discussion and Conclusion}
From the first experiment, the tractographic feature has the best mRS prediction accuracy and the lowest average absolute error compared to other first-order features.
The main reason is that the tractographic feature integrates volumetric-spatial information of the stroke lesion and the average diffusion information from a group of normal subjects that describes the potentially damaged regions impacted by the stroke lesion. These potentially damaged regions are formatted in the disruption matrix $D$ from \cref{eq:disruption}, and the weight vector $\vec{\gamma}$ from \cref{eq:weighted_tf} carries spatial and volumetric information of the stroke lesion to the tractographic feature $\vec{T}$.
In addition, it is worth noting that the volumetric-spatial feature is the same as the weight vector $\vec{\gamma}$ of the tractographic feature, and the mRS prediction performance of volumetric-spatial feature is improved by considering the average connectome information from a group of normal subjects.

The second experiment demonstrates that the tractographic feature also has better mRS prediction accuracy than the state-of-the-art feature \cite{maier2016predicting}. It should be noted that their approach requires ADC maps that are not necessarily always available, and using only the lesion shape information degrades the overall performance significantly in their approach. 
We also note that the tractographic feature is of much lower dimensions (116) compared to the state-of-the-art feature (1662).

\begin{figure*}[htbp]
    \centering
    \includegraphics[width=\textwidth]{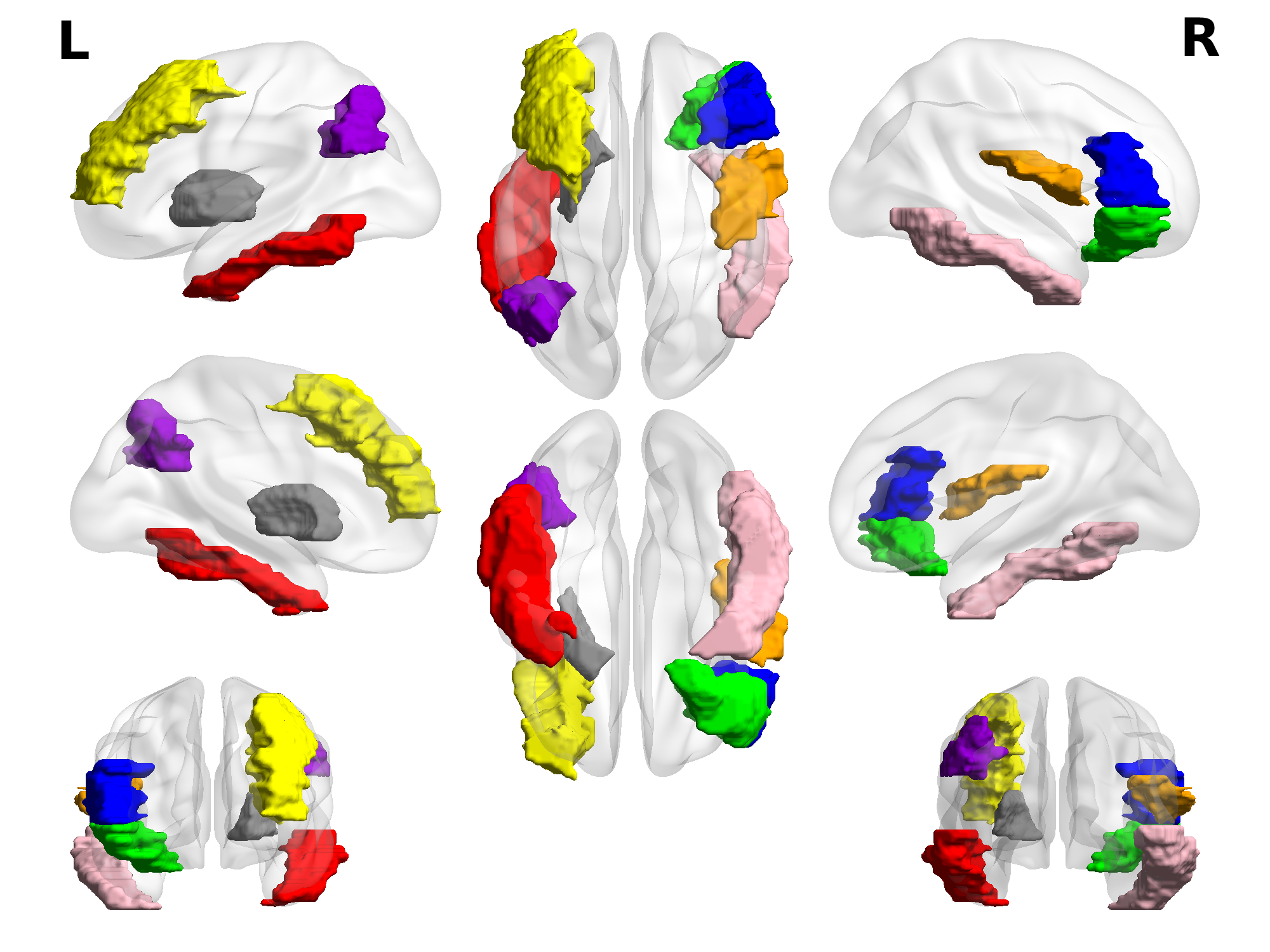}
    \caption{Selected tractographic feature from eight AAL regions including left (in red) and right (in pink) inferior temporal gyrus red, right Rolandic operculum (in orange), left middle frontal gyrus (in yellow), orbital part (in green) and triangular part (in blue) of right inferior frontal gyrus, left angular gyrus (in purple) and left putamen (in grey) after applying the recursive feature selection with cross-validation on the original tractographic features. These tractographic features are extracted from 37 ISLES 2017 training subjects. \textbf{Best viewed in color.}}
    \label{fig:tract_regions}
\end{figure*}

In both experiments, we apply the recursive feature selection with cross-validation on different types of features, and this procedure reduces one dimension of feature recursively until finding the best subset of the feature with the lowest mean absolute error. For the tractographic feature, this reduces the dimensionality from 116 to 8.
This selected tractographic feature comes from eight AAL regions shown in Fig.~\ref{fig:tract_regions}(left and right inferior temporal gyrus, right Rolandic operculum, left middle frontal gyrus, orbital part and triangular part of right inferior frontal gyrus, left angular gyrus and left putamen).

\begin{figure*}[htbp]
    \centering
    \includegraphics[width=\textwidth]{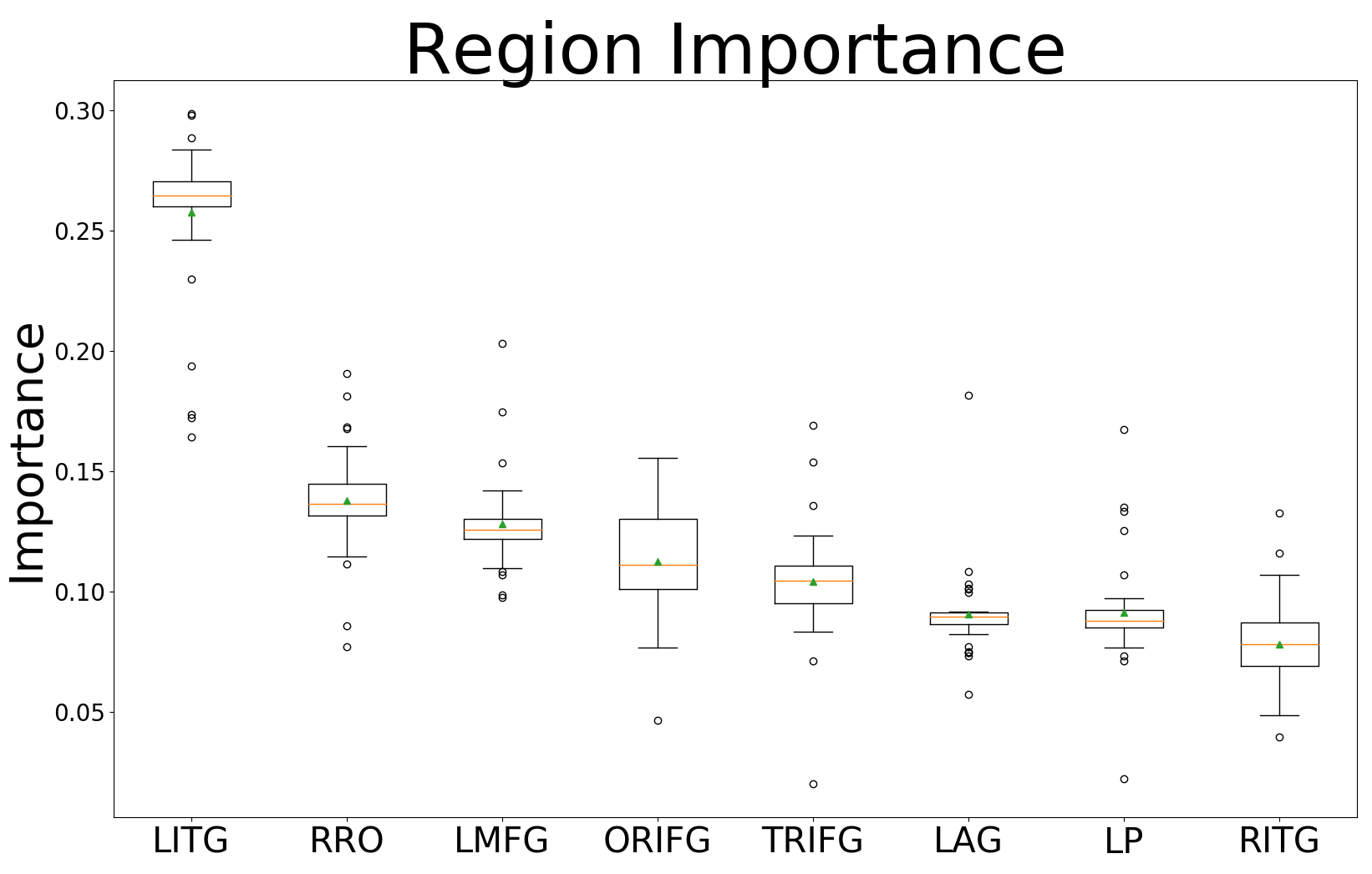}
    \caption{Region importance of eight selected AAL brain parcellation regions given by a random forest regressor with 300 trees whose maximum depth is 3. The average values are marked in the green triangles. Left inferior temporal gyrus (LITG) yields a higher mean importance (0.26) than right Rolandic operculum (RRO, 0.14), left middle frontal gyrus (LMFG, 0.13), orbital part (ORIFG, 0.11) and triangular part (TRIFG, 0.10) of right inferior frontal gyrus, left angular gyrus (LAG, 0.09), left putamen (LP, 0.09) and right inferior temporal gyrus (RITG, 0.08) within 37 ISLES 2017 training subjects on the task of predicting the mRS grades of stroke patients. \textbf{Best viewed in color.}}
    \label{fig:region_importance}
\end{figure*}

After feature selection, we use a random forest regressor to predict the mRS grades of stroke patients.
The random forest regressor gives the importance to each dimension within a given type of feature shown in Fig.~\ref{fig:region_importance}. 

\begin{figure*}[htbp!]
    \centering
    \begin{overpic}[width=0.32\textwidth]{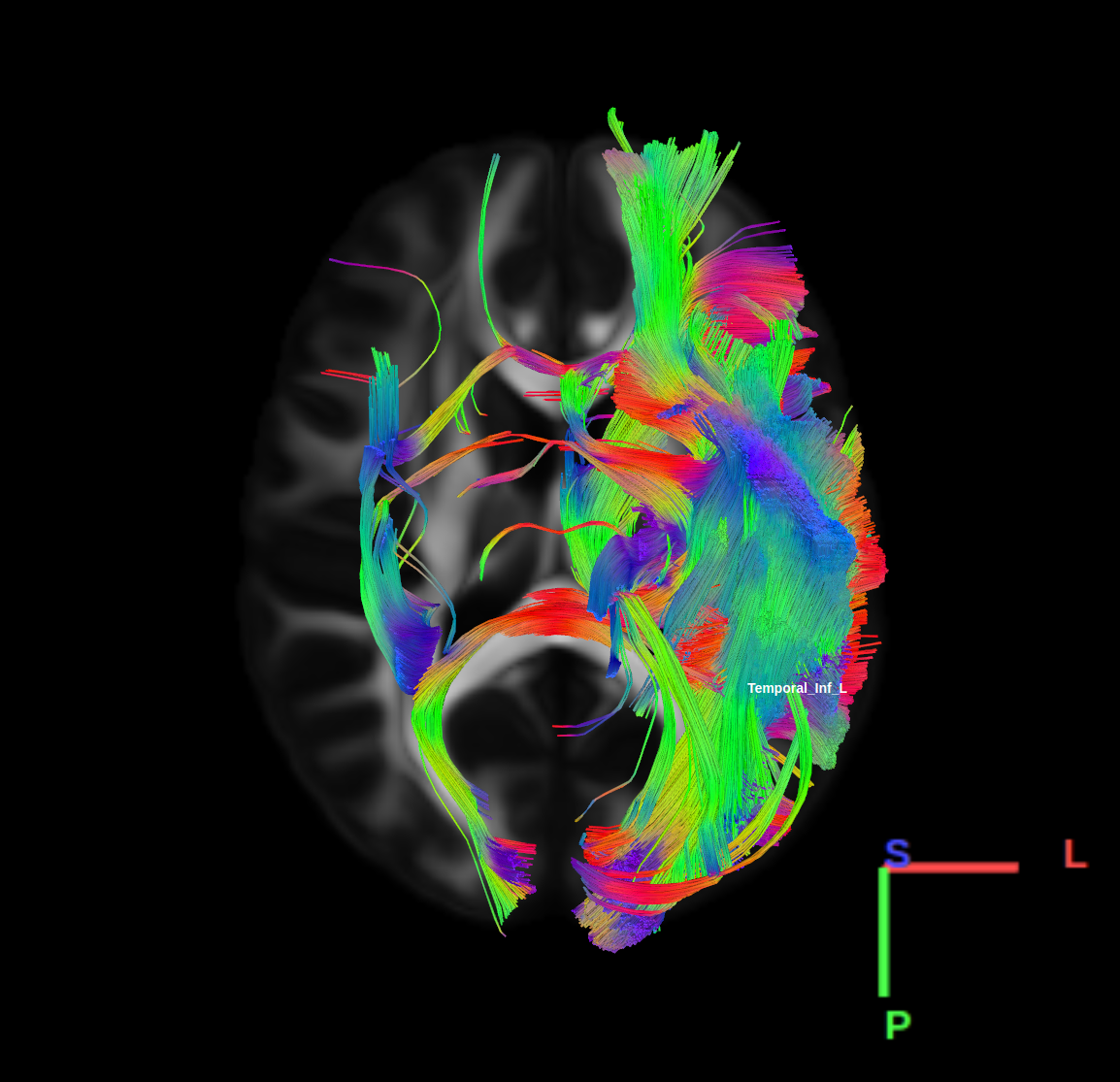}\end{overpic}
    \begin{overpic}[width=0.32\textwidth]{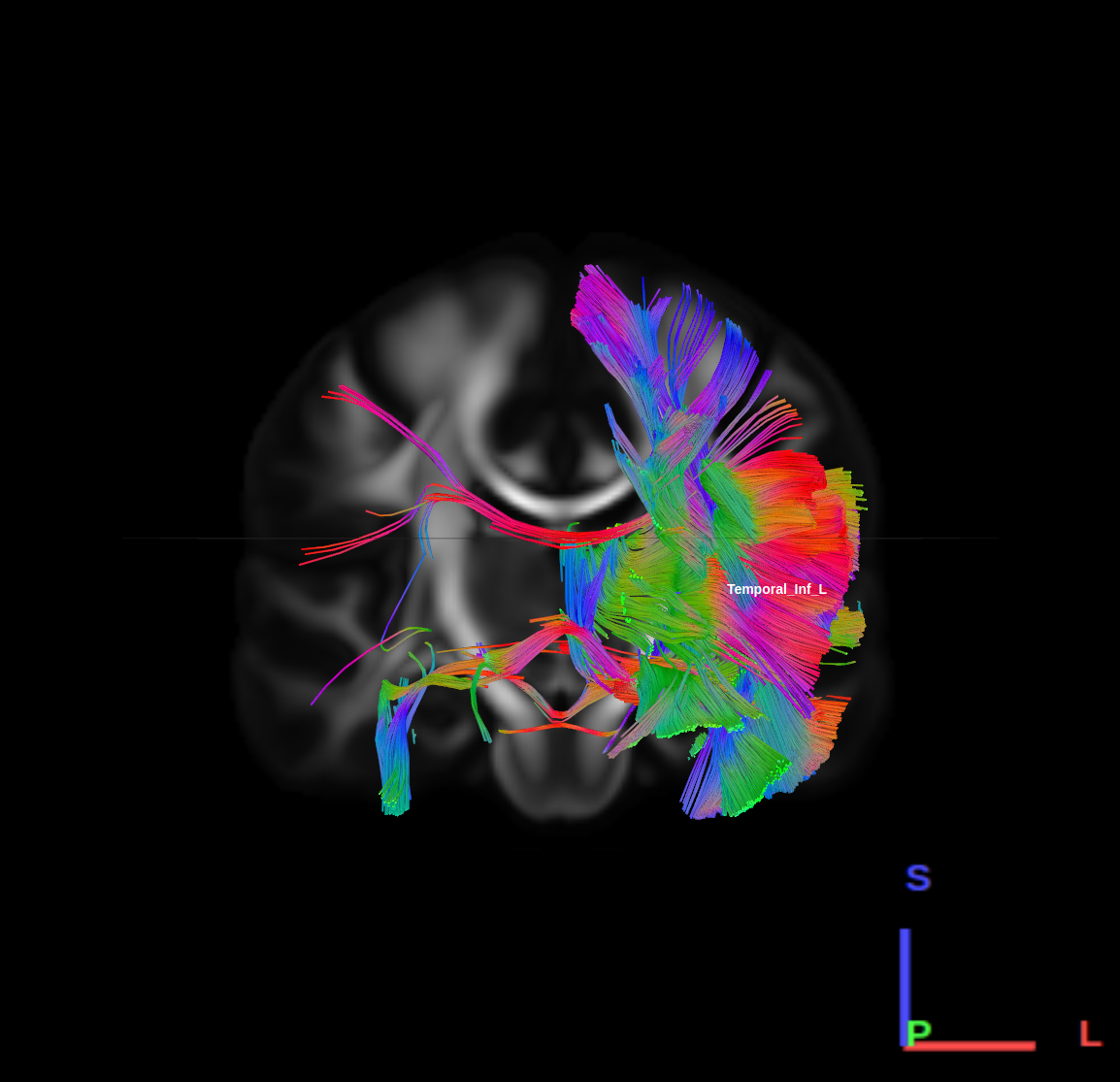}\end{overpic}
    \begin{overpic}[width=0.32\textwidth]{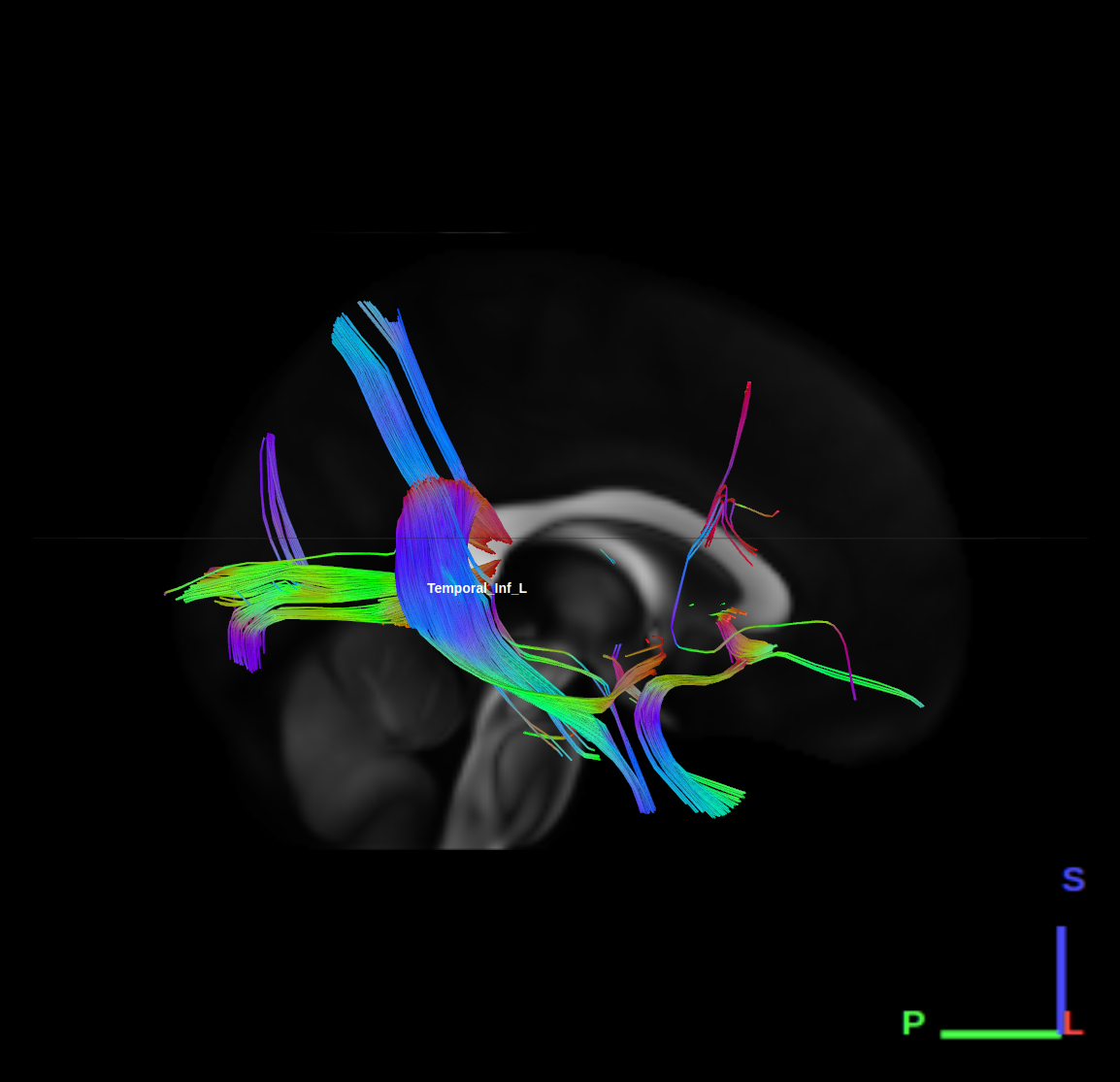}\end{overpic}\vspace{0.1mm}
    \begin{overpic}[width=0.32\textwidth]{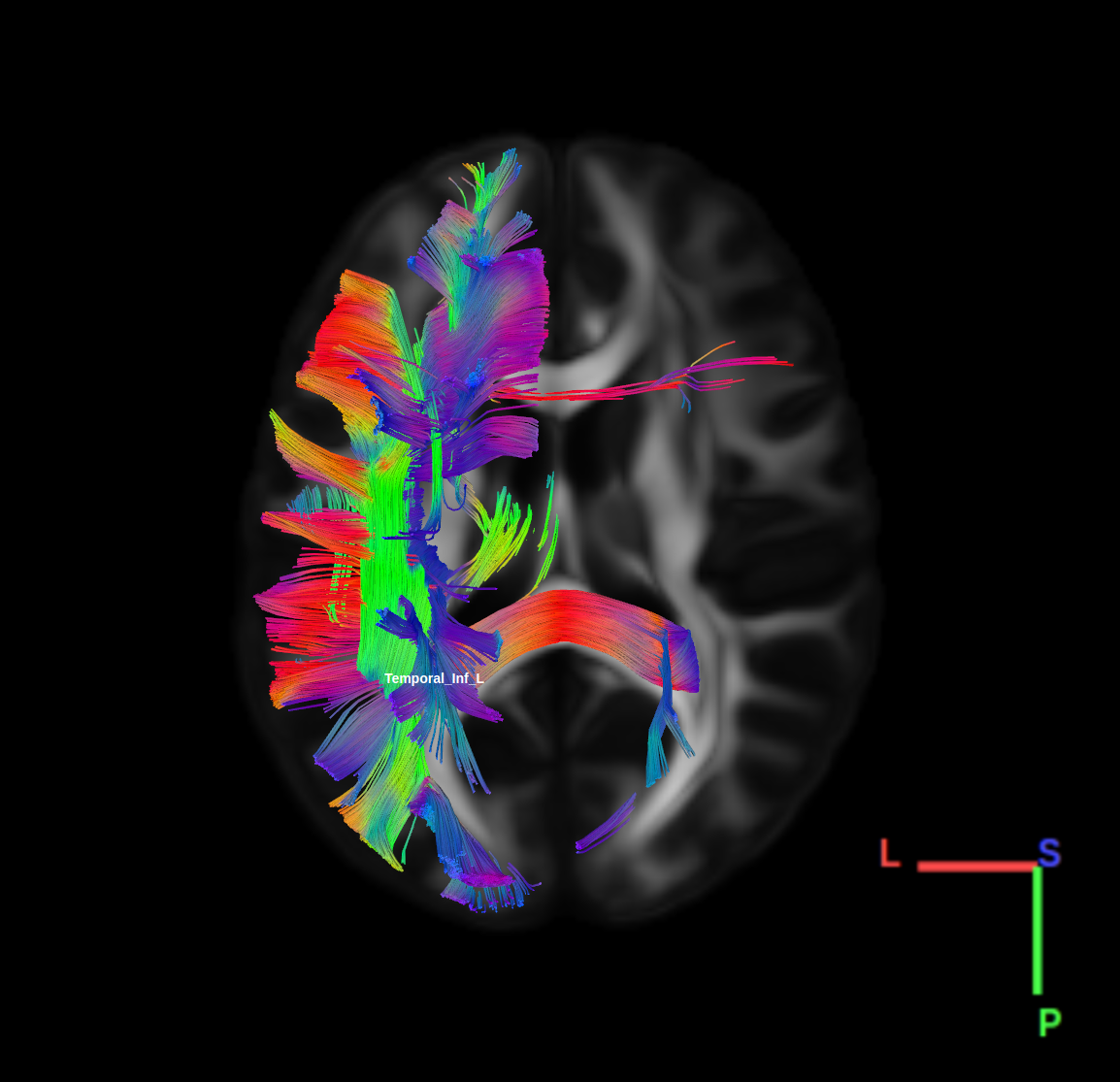}\put(5,3){\color{white}Axial View}\end{overpic}
    \begin{overpic}[width=0.32\textwidth]{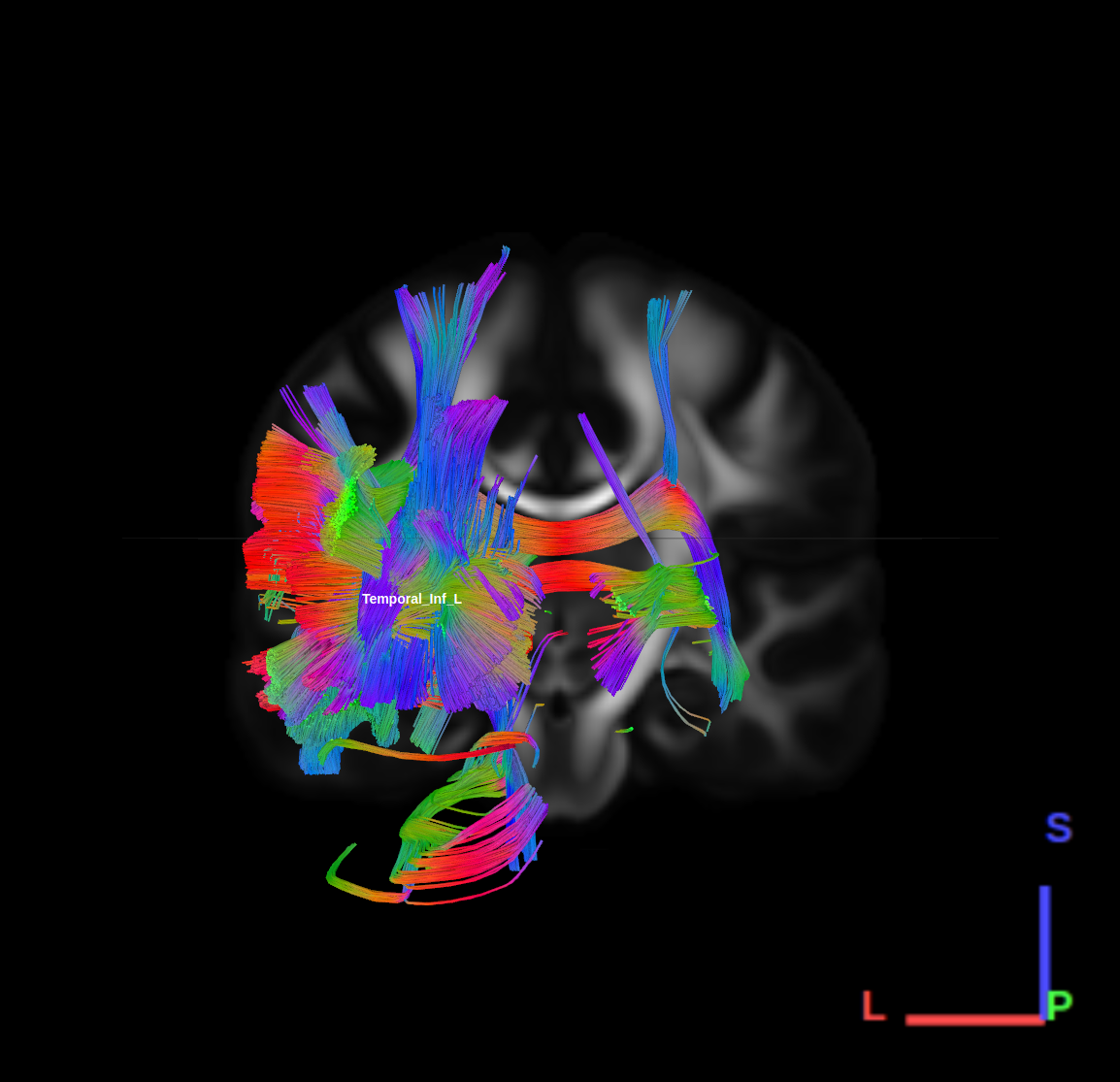}\put(5,3){\color{white}Coronal View}\end{overpic}
    \begin{overpic}[width=0.32\textwidth]{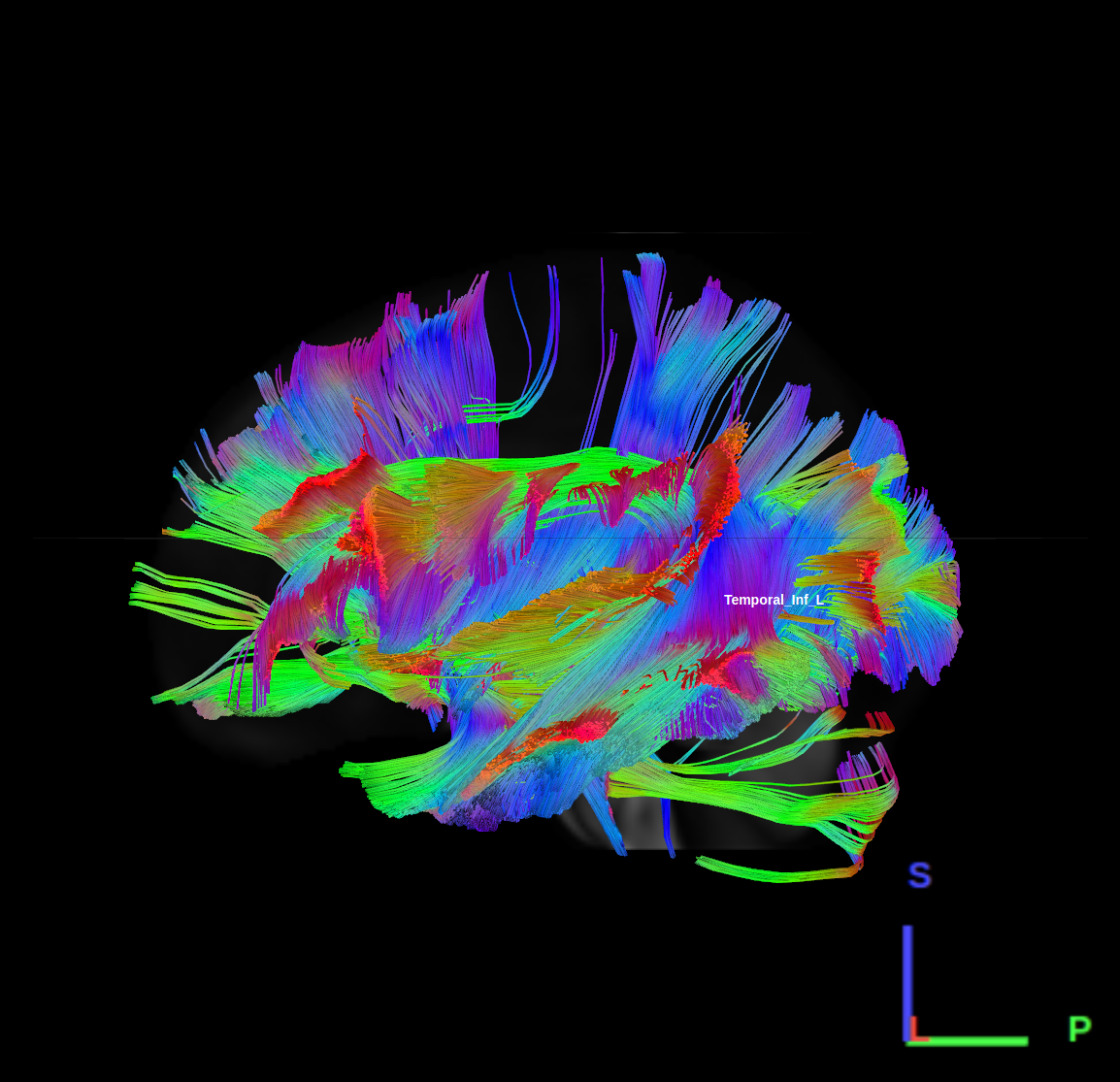}\put(5,3){\color{white}Sagittal View}\end{overpic}
    \caption{The fiber tracts passing through the left inferior temporal gyrus from the average connectome information of 1024 HCP subjects. We place a seed in each voxel inside the whole brain to find all possible tracts passing through the left inferior temporal gyrus. \textbf{Best viewed in color.}}
    \label{fig:til_tracts}
\end{figure*}

For the selected tractographic feature, left inferior temporal gyrus yields the highest average importance compared to the other seven regions within 37 ISLES 2017 training subjects on the task of predicting the mRS grades. 
The reasons left inferior temporal gyrus has the greatest effect on the mRS of stroke patients are (i) this region is important for language processing and speech production \cite{antonucci2008lexical}, and (ii) a large number of fiber tracts, passing through this region, goes across the splenium of the corpus callosum which connects the visual, parietal and auditory cortices \cite{hofer2006topography,knyazeva2013splenium} (See Fig.~\ref{fig:til_tracts}).

In conclusion, the paper presents for the first time the use of tractographic features for predicting the clinical outcome of stroke patients. The tractographic feature leads to promising mRS prediction results on ISLES 2017 dataset but needs to be further validated using a larger and representative independent dataset in order to rule out a potential methodical bias and over-fitting effects. The proposed tractographic feature has a potential to be improved if we build a disruption matrix from each HCP subject given the stroke lesion in MNI space and construct the average disruption matrix from these individual disruption matrices. 

\subsubsection{Limitation} The proposed tractographic feature cannot be generated if the stroke lesion is not located in the brain parcellation regions.

\section*{Acknowledgement}
This research was partially supported by a National Institutes of Health (NIH) award \# 5R01NS103774-03. We thank Oytun Ulutan for technical support, and Dr. Robby Nadler for writing assistance and language editing.

\bibliographystyle{splncs04}
\bibliography{mybibfile}
\end{document}